\documentclass[usenatbib]{mn2e}  
\usepackage{graphicx}
\usepackage{txfonts}

\usepackage{dcolumn}

\usepackage{natbib}

\DeclareMathAlphabet{\mathsc}{OT1}{cmr}{m}{sc}
\def\testbx{bx}%
\DeclareRobustCommand{\ion}[2]{%
\relax\ifmmode
\ifx\testbx\f@series
{\mathbf{#1\,\mathsc{#2}}}\else
{\mathrm{#1\,\mathsc{#2}}}\fi
\else\textup{#1\,{\mdseries\textsc{#2}}}%
\fi}

\newcommand{\HI}{\ion{H}{i}}
\newcommand{\CI}{\ion{C}{i}}
\newcommand{\CII}{\ion{C}{ii}}
\newcommand{\CIIs}{\ion{C}{ii}$^*$}

\newcommand{\MgII}{\ion{Mg}{ii}}

\newcommand{\OVI}{\ion{O}{vi}}
\def\lya{\ensuremath{{\rm Ly}\alpha}}

\def\lyb{\ensuremath{{\rm Ly}\beta}}
\def\kms{km\,s$^{-1}$}
\def\psavz{${<{{\dot{\psi_{*}}}}(z)>}$}

\begin{document}

\title[\lya\ emission from DLAs]{
The Lyman-$\alpha$ emission of high-$z$ damped Lyman-$\alpha$ systems
}
\author[Rahmani et al. ]{H. Rahmani$^{1}$\thanks{E-mail: hadi@iucaa.ernet.in}
R. Srianand$^{1}$, P. Noterdaeme$^{2}$ \& P. Petitjean$^{3}$
\\
$^{1}$ Inter-University Centre for Astronomy and Astrophysics, Post Bag 4,  Ganeshkhind, Pune 411\,007, India \\
$^{2}$ Departamento de Astronom\'ia, Universidad de Chile, Casilla 36-D, Las
Condes, Santiago, Chile\\
$^{3}$ Universit\'e Paris 6, UMR 7095, Institut d'Astrophysique de Paris-CNRS, 98bis Boulevard Arago, 75014 Paris, France \\
}
\date{Accepted. Received; in original form }
\pagerange{\pageref{firstpage}--\pageref{lastpage}} \pubyear{2009}
\maketitle
\label{firstpage}


\begin {abstract}
{
Using a spectral stacking technique we searched for the average \lya\ 
emission from  high-$z$ Damped  \lya\ (DLA) galaxies detected in the 
Sloan Digital Sky Survey QSO spectra.
We used a sample of 341 DLAs of mean redshift $<z>$~=~2.86 and 
log~$N$(H~{\sc i})$\ge$20.62 to place a 3$\sigma$ upper limit of 
3.0$\times10^{-18}$~erg~s$^{-1}$~cm$^{-2}$ on the \lya\  flux 
emitted within $\sim$1.5~arcsec (or 12~kpc) from the QSO line of sight. 
This corresponds to an average \lya\ luminosity  of $\le2\times 10^{41}$ 
erg s$^{-1}$  or
0.03 $L_\star$(\lya). 
This limit is deeper than the limit of most surveys for faint \lya\ 
emitters. The lack of \lya\ emission in DLAs is consistent with the in situ 
star formation, for a given $N$(H~{\sc i}), being less efficient than 
what is seen in local galaxies.
Thus, the overall DLA population seems to originate from the low 
luminosity end of the high redshift \lya\ emitting galaxies and/or to be 
located far away 
from the 
star forming regions. The latter may well be true since we detect strong O~{\sc vi} absorption 
in the stacked spectrum, indicating that DLAs are associated with a highly ionized phase possibly
the relics of galactic winds and/or originating from cold accretion flows.  
%
%
We find the contribution of  DLA galaxies to the global star formation rate density 
to be  comparatively lower than that of Lyman Break Galaxies. 
}
\end{abstract}



%
\section{introduction}
Damped \lya\ (DLA) systems, detected in absorption in the spectra 
of background quasars are characterized by large neutral hydrogen column 
densities ($N(\HI) \ge 2\times 10^{20}$~cm$^{-2}$) and represent the main 
reservoir of H~{\sc i} at high-$z$ \citep[see][Fig. 14]{Noterdaeme09dla}. 
The presence of associated heavy elements, the evolution with redshift of the 
mass density in DLAs, a signature of gas consumption 
via star formation  and the detectability of  
DLAs over a wide range of redshift make them the appropriate 
laboratories for studying the cosmological evolution of star formation
activity in a luminosity unbiased way. 

Though observational studies of DLAs have been pursued over 25 years, 
one of the most important questions that remain unanswered yet
is the connection between DLAs and star-forming galaxies.
Measuring the star formation rate (SFR)
in DLA-galaxies is very important as DLAs could provide substantial 
contributions to the global SFR  density at high 
redshifts \citep{Wolfe03b,Srianand05}. 

At low and intermediate redshifts ($z < 1$), galaxy counterparts of 
DLAs have been identified in a number of cases 
\citep{Burbidge96,LeBrun97,Rao03}. 
They are usually low surface brightness dwarf galaxies 
\citep[see][]{Rao03}.
On the other hand, 
searches for the direct emission of 
the high redshift (i.e $z>1$) DLA galaxies have
resulted in a number of non-detections 
 \citep{Lowenthal95,Bunker99,Colbert02,Kulkarni06,Christensen07}
and only a few cases have been spectroscopically confirmed  
through detection of \lya\ emission
\citep{Warren96,Moller98b,Moller02,Moller04,Heinmuller06,Fynbo10}. 
%
%
The difficulty is mainly attributed to the faint nature of the DLA
galaxies and their apparent closeness to the bright background QSOs.

%
One indirect way to address this question is to 
study the relative populations of different rotational levels 
of molecular hydrogen (H$_2$) together with that of fine-structure levels of
\CI\ and \CII. The ambient UV flux and therefore the in-situ SFR
can thus been derived 
\citep[e.g.][]{Hirashita05,Srianand05,Noterdaeme07} in a few of the
$\sim10$\% of DLAs that show detectable amounts of H$_2$ 
\citep{Petitjean00,Ledoux03,Noterdaeme08}. The rotational
excitation seen in these H$_2$ bearing DLAs is consistent with an
ambient UV radiation field similar to or higher than
the Galactic one \citep{Srianand05}. 
This may not be representative of the overall population however.

Another indirect approach is to derive the SFR through the
cooling rate inferred from the \CIIs$\lambda1335$\ absorption lines  
as suggested by \citet{Wolfe03a}. They have estimated an average 
SFR per unit area  of 10$^{-2.2}$ M$_\odot$ yr$^{-1}$ kpc$^{-2}$
or 10$^{-1.3}$ M$_\odot$ yr$^{-1}$ kpc$^{-2}$ if 
DLAs arise in cold neutral medium (CNM) or
warm neutral medium (WNM) respectively. 
However, appreciable contribution of \CIIs\ absorption from the warm ionised medium (WIM) 
as observed in the diffuse gas of the Milky Way \citep{Lehner04} will 
make the inferred SFR based on \CIIs
absorption lines unreliable. 

Thanks to the availability of a very large number of moderate resolution QSO
spectra in the SDSS data base, it is possible to
directly detect emission from the galaxy counterparts within 1.5 arc sec
to the line of sight \citep{Noterdaeme10o3} and/or to measure the average
SFR in different types of QSO absorption systems by
detecting nebular emission in the stacked spectra \citep{Wild07,Noterdaeme10o3,Menard09}. 
Here, we tried to measure the average
\lya\ emission from DLAs using spectral stacking techniques.
We describe our DLA sample in Section \ref{DLAsample}, 
discuss the stacking technique we use and the results 
in Section~\ref{stack}. We discuss the \OVI\ detections in
Section~\ref{OVI} and contribution of DLAs to 
%
%
global SFR density in
Section~\ref{SFR_G}. Our conclusions are presented 
in Section~\ref{conclude}.  Throughout this paper we assume a standard flat
$\rm{\Lambda CDM}$ universe with $\rm{H_{0}}=71~\rm{km\,s^{-1}\,Mpc^{-1}}$, 
$\rm{\Omega_{M}}=0.26$, $\rm{\Omega_{\Lambda}}=0.74$.
%
\section{DLA Sample} \label{DLAsample}
The sample of DLAs used here mainly comes from an automatic search 
for DLAs in the Sloan Digital 
Sky Survey II, Data Release 7 \citep{Noterdaeme09dla}. 
This sample contains 1426 absorbers at $2.15 < z < 5.2$ with 
$\log N($\HI$)\ge20$, out of which 937 systems have $\log N($\HI$)\ge 20.3$. 
We also use additional DLAs  found by \citet{Prochaska09s}. 
In total, there are 914 DLAs with redshift measurements 
based on metal absorption lines. We notice that the poor background 
subtraction leaves some spikes in the locations of strong 
sky lines. To avoid any systematics due to these features 
we have excluded systems located in regions of strong sky lines 
\footnote{The wavelength range affected by sky lines
are 4040-4060{\AA}, 4340-4380 {\AA}, 5170-5230 {\AA}, 5450-5475 {\AA}
and $\ge$ 5560{\AA}.}.

In order to establish the presence or absence of any \lya\ emission,
we need the bottom of the DLA absorption profile to be defined by
enough pixels with zero flux. We will use only those DLAs for which the 
core (with zero flux) extends over at least three FWHM elements. 
Since the spectral resolution of SDSS spectra is about $R=2000$, the 
rest FWHM
of an unresolved \lya\ emission 
is $\sim 0.61$~{\AA}.
The above condition translates to a  lower limit on the
H~{\sc i} column density of $\log$ $N(\HI)=20.62$.
%

By 
visual inspections of the individual spectra,
we rejected systems showing a double absorption feature in either 
their \lyb\ or metal absorption profiles. Indeed some of these systems may not
be real DLAs as the high \lya\ equivalent width may be the result
of the blend of several components.
We also rejected proximate DLAs located within 5000~km~s$^{-1}$ from the QSO redshift and ensured  
that there is no contamination of the sample by broad absorption lines. 
After applying all these conditions we are left with 341 systems
with log~$N$(H~{\sc i})$\ge$20.62 that forms our sample to be used in the stacking
exercise.  
\section{Spectral Stacking} \label{stack}

%
\begin{figure*} 
\centering
\hbox{
\includegraphics[width=0.33\hsize,bb=30 47 550 715,clip=]{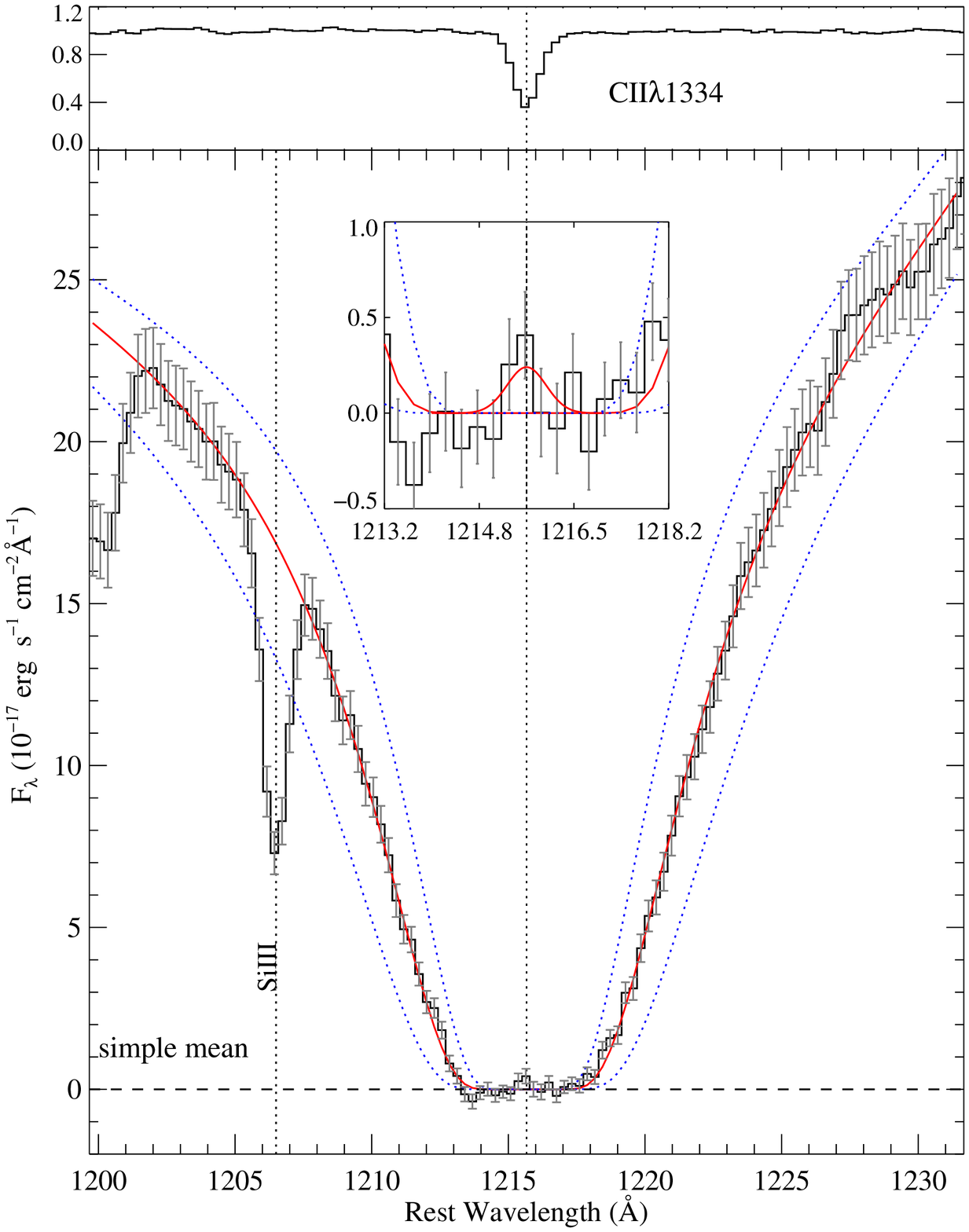}
\includegraphics[width=0.33\hsize,bb=30 47 550 715,clip=]{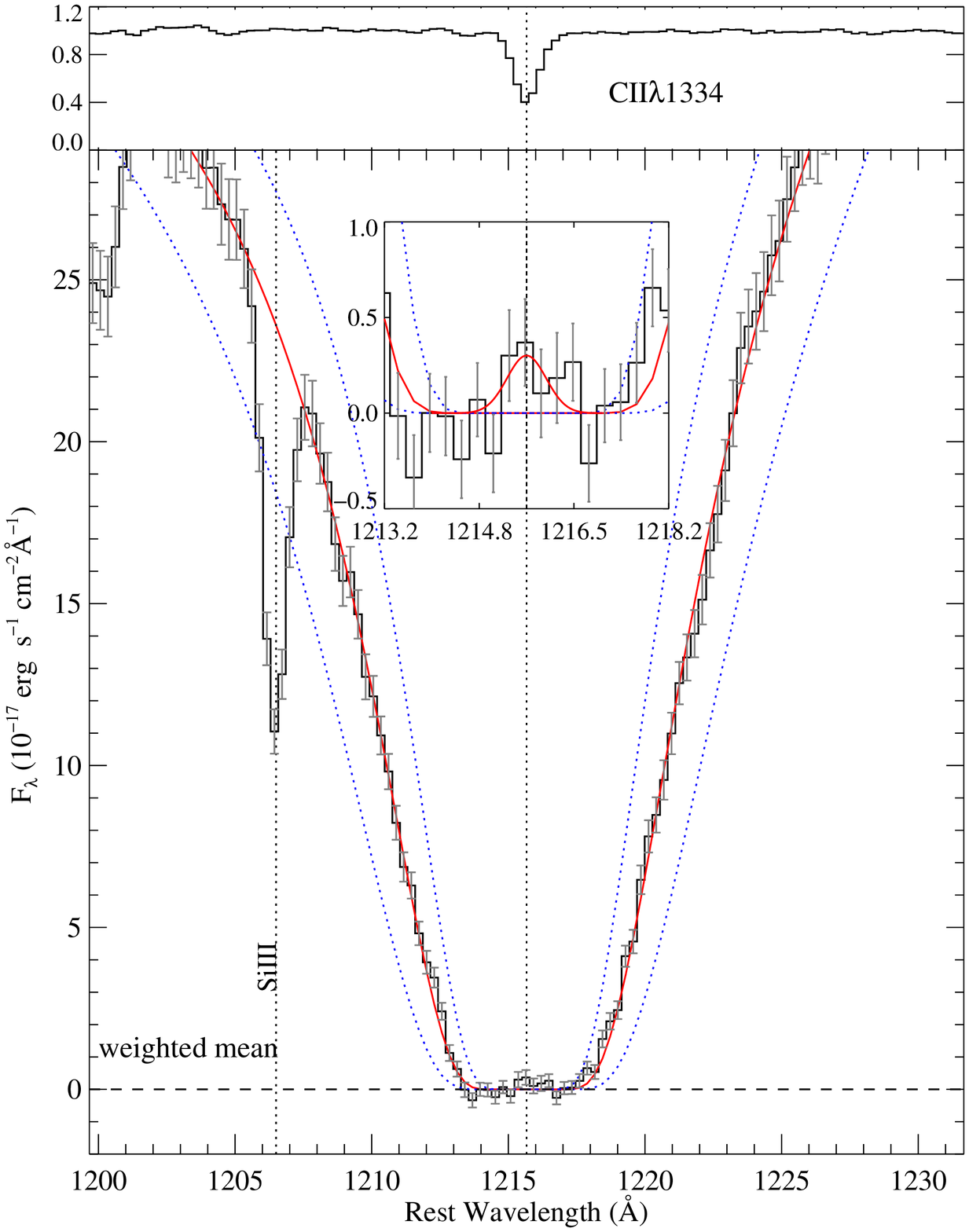}
\includegraphics[width=0.33\hsize,bb=30 47 550 715,clip=]{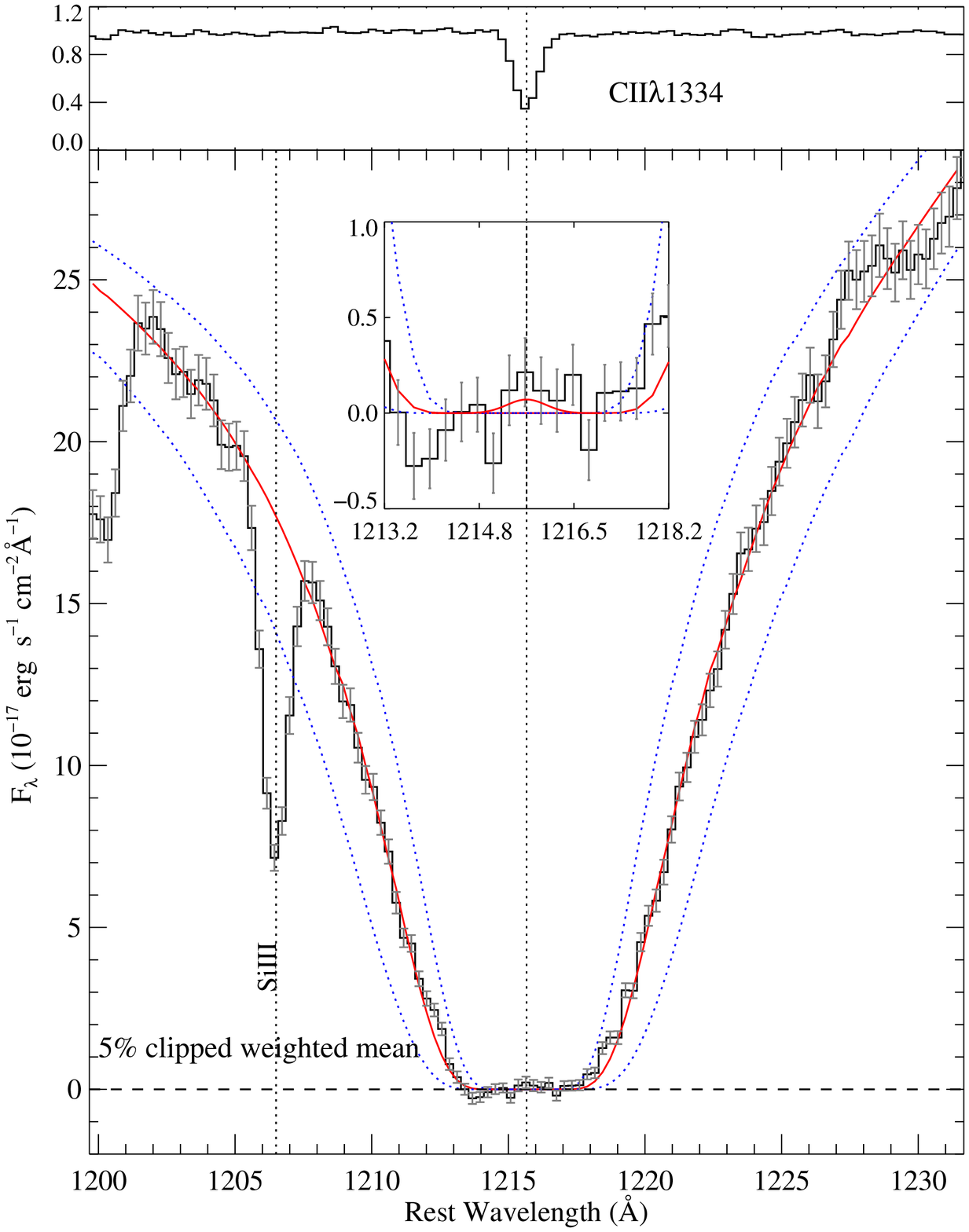}
}
\caption{Stacked spectrum of DLAs obtained from simple mean, weighted mean, and weighted mean 
with 5\% clipping are shown in, respectively, the left, middle and right panels. The red continuous 
curves are the stacked spectrum of the DLA synthetic absorption profiles. 
The dashed blue curves are the synthetic profiles obtained using the lower and 
upper limits on $N(\HI)$ as measured in each DLA. A gaussian with $FWHM=200$~km~s$^{-1}$ is fitted 
in the bottom of the absorption trough and is shown by 
a red continuous curve in the inset. {\it Top panels}: profile of
C~{\sc ii}$\lambda$1334 line to give an idea of the velocity spread of
neutral gas.
} 
\label{stacked}
\end{figure*}
%

The observed SDSS spectra 
were first shifted to the rest frame of the DLA 
conserving the flux per unit wavelength interval. 
The residual flux level in
the core of the DLA absorption profile generally shows some non-zero
off-set probably related to sky subtraction errors. We correct this off-set using the median flux
in the absorption  trough before stacking. While this exercise makes our 
stacked spectrum insensitive to  the presence of continuum light from the DLA galaxy, 
it will not affect the detectablity of the \lya\ line.

The flux in each pixel of the combined spectrum  is calculated 
by averaging the fluxes of all spectra at this position. Arithmetic mean and  
weighted mean are both used.
In the case of weighted mean the i-band signal-to-noise ratio (SNR) 
is used as the weighting factor for each spectrum. 
We calculate the uncertainty in each pixel of the combined spectrum as the standard 
deviation around the mean value.
We obtain a third average spectrum from the weighted mean of the values in 
each pixel after rejecting the 5\% extreme values on either positive and 
negative sides.

The stacked spectra obtained using simple mean, weighted mean, and weighted mean with clipping 
together with associated errors are shown in Fig.~\ref{stacked}.
We also show the stacked synthetic absorption DLA profiles in the corresponding samples 
as red continuous curves. The dashed blue curves are the synthetic 
profiles obtained using the lower and upper limits on $N(\HI)$ obtained in individual systems.
To calculate the \lya\ emission flux, in each stacked spectrum, we restrict 
ourselves to the wavelength range covered by the core of synthetic spectrum
obtained from the lower limits on $N$(H{\sc i}) (i.e inner 
dotted profiles in Fig.~\ref{stacked}).
%
We obtain the limit on the \lya\ emission line flux by integrating the observed flux over
a gaussian function
with $\rm{FWHM}$ of 200~km~s$^{-1}$.
This value corresponds to the mean velocity widths of low
ionization lines in DLAs \citep{Ledoux06a} convolved with the SDSS  instrumental broadening.
{The observed FWHM of the low ionization lines in our composite 
spectrum is consistent with this value}.
%
%
\begin{table} 
 \centering
  \caption{\lya\ measurement in the stacked spectrum. }
  \begin{tabular}{lccc}
  \hline
\hline
   Stacking method  &${F({\rm \lya})}^ a$ &${F_{3\sigma}~({\rm \lya})}^ b$&   ${L_{3\sigma}~({\rm \lya})}^ b$\\
  \hline
  simple mean                 &0.21$\pm$0.13  &0.39  &  28.3  \\
  weighted mean               &0.26$\pm$0.13  &0.39  &  28.3 \\
  weighted mean with clipping &0.13$\pm$0.10  &0.30  &  21.8 \\
\hline
\\
\multicolumn{4}{l}{$^ a$ in units of  10$^{-17}$ ergs s$^{-1}$ cm$^{-2}$}\\
\multicolumn{4}{l}{$^b$  in units of  10$^{40}$ergs~s$^{-1}$}\\
\end{tabular}
\label{table1}
\end{table}

The \lya\ flux measurements are summarised in Table~\ref{table1} and the gaussian fits are 
shown in the insets of Fig.~\ref{stacked}. 
We do see marginal \lya\ flux at the expected position at a $\le$ 2$\sigma$ level in the case of 
simple and weighted mean. In the case of clipped mean the bottom of the \lya\ absorption
profile is consistent with no \lya\ emission.
%
%
We estimate the 3$\sigma$ upper limit ({${F_{3\sigma}~({\rm \lya})}$} given in 
Table~\ref{table1}) by using the gaussian fitting errors. As expected, the 
3$\sigma$ limit from the clipped mean is smaller than that from the other 
two methods. As this is the case with less contamination by outliers, we use 
the upper limit from the clipped mean composite spectrum,
${F_{3\sigma}~({\rm \lya})} < 3.0 \times 10^{-18}$~erg s$^{-1}$ cm$^{-2}$, 
for all further discussions. 
{For the SDSS fibre diameter, 3 arcsec,
this limit translates to a limiting 1$\sigma$ surface brightness limit 
of 1.4$\times10^{-19}$ ergs s$^{-1}$ cm$^{-2}$ arcsec$^{-2}$. This 
is only 1.7 times higher than that achieved in the very deep
long slit spectroscopic observations by \citet{Rauch08}.}

The flux limit  we have achieved in the stacked spectrum is almost one order 
of magnitude smaller than  the flux measured in the case of direct 
detections of \lya\ emission from DLA galaxies \citep[see][]{Moller04,Fynbo10} and the typical flux limits reported for individual
measurements \citep[see][]{Christensen07}. 
In addition the flux limit we have reached is much deeper than the one reached in typical
blind narrow band searches for \lya -emitters 
\citep[see][]{Ouchi08}
and comparable to the deepest limit achieved in the VVDS-LAE survey by \citet{Cassata10}.

For the mean redshift of our sample, $<z>=2.86$, 
our limiting flux corresponds to a \lya\ luminosity of 2$\times 10^{41}$ erg s$^{-1}$ (see Table~\ref{table1})
or $\sim 0.03 L^{\star}({\rm \lya})$ if we use the Schechter function
parameter $L^{\star}=5.8\times 10^{42} \rm{erg~s^{-1}}$ 
from the narrow band \lya\ search by \citet{Ouchi08}. 
%
Thus it appears that on an average, DLAs at $z\sim$2.8 trace the faint end 
luminosity of \lya\ emitters.

\section{star formation} \label{SFR_G}

%
\begin{figure} 
\centering
\includegraphics[width=0.95\hsize,bb=65 170 485 575]{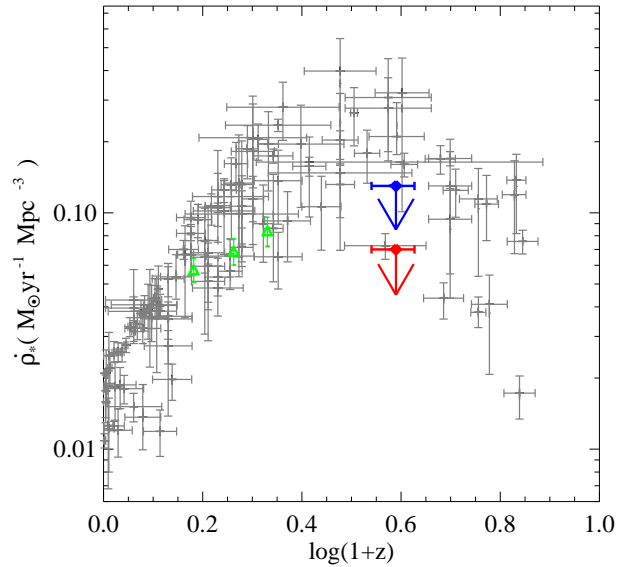}
\caption{Evolution of the cosmic star formation rate density with redshift. The 
black points with error bars are taken from \citet{Hopkins06}. 
The red point refers to the $3\sigma$ upper limit on the contribution 
of DLA galaxies with log~$N(\HI) \ge 20.62$ as measured in this work. 
The blue point is the same assuming all DLAs with log~$N(\HI) \ge 20.30$ 
contribute the same.
The green points show the contribution of \MgII\ systems  estimated by \citet{Menard09}.}
\label{hopkins}
\end{figure}

%
We estimate the average SFR in DLAs assuming  that
\lya\ photons mainly originate
from H~{\sc ii} regions around massive stars and
case B recombination  \citep{Osterbrock89book}. 
Then the \lya\ luminosity 
is related to the SFR (${\dot M}_{\rm SF}$) by,
\begin{equation}
L_{{\rm Ly\alpha}} = 0.68 h \nu_{\alpha} ( 1 - f_{\rm esc} ) {N}_\gamma {\dot M}_{\rm SF} .
\label{eqn_lyman}
\end{equation}
Here, $h\nu_{\alpha}=10.2$~eV and $f_{\rm esc}$
are, respectively, the energy of a Lyman-$\alpha$ photon and the 
escape fraction of Lyman continuum photons.
We use $f_{\rm esc}\simeq 0.1$ as measured from
stacked spectra of high-$z$ LBGs
\citep{Shapley06}.
Further, ${N}_\gamma$ is the number of ionizing photon s released
per baryon of star formation. 
This is mainly a function of the metallicity and the stellar initial mass function
 \citep[see Table 1 of][] {Samui07}. As high redshift DLAs have generally 
low metallicities, $[Z /Z_{\odot}] \sim -1.5$ \citep[see Fig 3 of][]{Noterdaeme08}, 
 we interpolate values of  Table 1 of \citet []{Samui07} to have 
$ {N}_\gamma=9870$ corresponding to  $ [Z /Z_{\odot}] = -1.5$
and a Salpeter initial mass function with $\alpha=2.35$.
\lya\ photons are sensitive to different kinds  of radiation
transport effects including resonant scattering (within the ISM 
as well as the intergalactic medium (IGM)) and attenuation via dust. 
Hence, the observed \lya\  luminosity 
can be related to the emitted luminosity by
\begin{equation}
L_{\rm Ly\alpha}^{\rm obs} = f_{\rm esc}^{\rm Ly\alpha} L_{\rm Ly\alpha},
\label{LaL}
\end{equation}
where $f_{\rm esc}^{\rm Ly\alpha}$ is the escape probability of the \lya\ photons. 
It is well known that only a small fraction of LBGs are \lya\ emitters and
even in these cases only a small fraction of \lya\ photons escape the galaxy
\citep[see][]{Kornei10}.
Taking this fact into account it is appropriate to use a volumetric $f_{\rm esc}^{\rm Ly\alpha}$
of 0.05 as estimated by \citet{Hayes10} for high-$z$ galaxies. With these assumptions, the 3$\sigma$
upper limit we have obtained above corresponds to a SFR of $\le  1.2$ M$_\odot$ yr$^{-1}$. 
For comparison, this limiting value is higher than the
average SFR of 0.1$-$0.5~M$_\odot$~yr$^{-1}$ measured in low-$z$ Mg~{\sc ii} and Ca~{\sc ii} systems
\citep{Wild07,Noterdaeme10o3,Menard09}.

The SFR per unit comoving volume for DLAs is given by
\begin{equation}
{{\dot{\rho_{*}}}}(z)={<{{\dot{\psi_{*}}}}(z)>}(A/A_{\rm p})d{\cal N}/dX
\label{rhosfr}
\end{equation}
\noindent where {\psavz} is the average SFR per unit
physical area at $z$, $A$ is the
average physical cross-sectional area, $A_{\rm p}$ is the average projection of 
$A$ on the plane of the sky, and  $d{\cal N}/dX$ is the  incidence of DLAs per 
unit absorption distance interval.
%
We convert the star formation rate ($\dot{M}_{\rm SF}$) to 
{\psavz} by assuming the typical size of DLAs to be similar to 
that of faint \lya\ emitters,  $R=10$ kpc. 
%

%
%
We estimate $d{\cal N}/dX$ to be $9.95\times 10^{-6} ~\rm{Mpc^{-1}}$
and  $\dot{\rho}_*\le  0.07$ M$_\odot$ yr$^{-1}$ Mpc$^{-3}$
for DLAs with log~$N(\HI) \ge 20.62$ at $<z> = 2.86$ assuming
$(A/A_{\rm p})$ = 2.
If we assume the upper limit on the \lya\ flux is also valid
for all DLAs with log $N$(H~{\sc i})$\ge$20.3, then we get 
$\dot{\rho}_*\le  0.13$ M$_\odot$ yr$^{-1}$ Mpc$^{-3}$. 
It is to be remembered that
increasing $R$ or $f_{\rm esc}^{\rm Ly\alpha}$ will further decrease $\dot{\rho}_*$.
In Fig.~\ref{hopkins} we compare these values with the extinction corrected 
comoving star formation rate density measurements summarized in \citet{Hopkins06}. 
The conservative upper limits from our study  are comparatively  lower than the 
measurements based on LBGs. 
%
%
\section{High-ionization gas probed by \OVI\ absorption}\label{OVI}
\begin{figure} 
\centering
\includegraphics[width=0.9\hsize,bb=19 75 580 600,clip=]{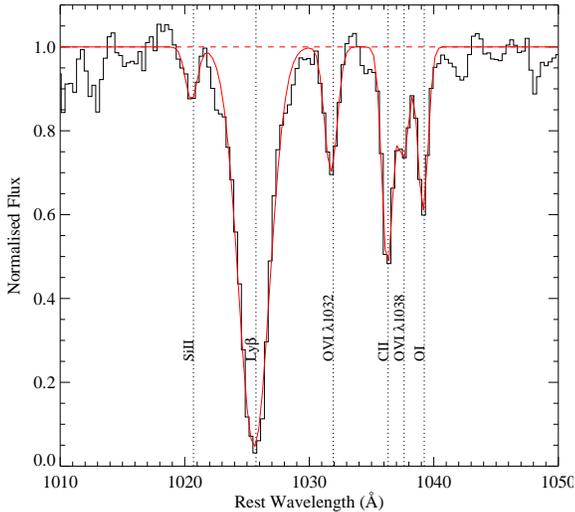}
\caption{Wavelength range of the SDSS composite spectrum encompassing the \OVI\ absorptions.
Gaussian fits to some of the absorption lines are also overplotted}
\label{figovi}
\end{figure}

If DLAs are associated with star forming regions, they should be associated
with hot gas related to galactic outflows and/or infall of primordial gas. 
Indeed, numerical simulations of cosmological structure formation 
\citep[e.g][]{Dave01}
%
predict that gas falling                                                      
onto diffuse large-scale structures will be accelerated to supersonic 
speeds and become shock-heated to temperatures of 10$^5$ to 10$^7$~K, creating 
a phase known as the warm-hot IGM. Similarly a sustained  
star formation activity in galaxies can lead to super novae driven outflows
of hot gas.
Observations of high-$z$ Lyman Break Galaxies (LBGs) frequently show galactic scale superwinds
and the mass outflow rate may be intimately related to the star formation
rate \citep{Pettini01lbg}. This hot gas can be traced by \OVI\ absorption.


\citet{Fox07b} have detected \OVI\ in at least 34\% of the DLAs and suggested
that \OVI\ could be present in 100\% of them.
%
We confirm this finding by detecting the \OVI\ doublet (with rest 
equivalent width $W_{\rm r}$(O~{\sc iv}$\lambda$1032)~=~0.4\AA) 
in our composite spectrum (see Fig.~\ref{figovi}). The average deconvolved velocity 
width of the \OVI\ lines ($\sim$275 \kms) is systematically higher than that 
of C~{\sc iv} ($\sim$250 \kms), Si~{\sc iv} ($\sim$205 \kms ) and 
singly ionization lines ($\sim$120 \kms). The velocity range spanned by the hot phase is therefore
a factor of two larger than that of the neutral phase.
This suggests a connection between actively star forming regions with outflowing and/or 
infalling hot gas and DLAs.

\section{Summary and discussion} \label{conclude}

Using a sample of 341 DLAs with log~$N$(H~{\sc i})$\ge$20.62 
we place a 
3$\sigma$ upper limit of 3.0$\times10^{-18}$~erg~s$^{-1}$~cm$^{-2}$ on 
the \lya\ flux emitted at the redshift of the DLA within $\sim$1.5~arcsec (or 12 kpc at
$z$$\sim$2.86) to the QSO line-of-sight.
This corresponds to an average \lya\ luminosity of
$\le2\times 10^{41}$ erg s$^{-1}$ or an object 
with a luminosity of 0.03~$L_\star$(\lya). Thus, the typical
DLA population seems to originate from the low luminosity end of
the high redshift \lya\ emitting galaxies.

At low redshift the star forming galaxies obey the Kennicutt \& Schmidt law,
\begin{eqnarray}
<\dot{\psi_\star}>_\perp & = & 0 ~{\rm when} ~
N_\perp < N_\perp^{\rm crit}\nonumber\\
                         & = &  K (N_\perp/N_\perp^{\rm c})^\beta,~~~ N_\perp\ge N_\perp^{\rm crit}.
\label{sklaw}
\end{eqnarray}
\citet{Kennicutt98, Kennicutt98b} have found $K=(2.5\pm0.5)\times10^{-4}$~M$\odot$~yr$^{-1}$~kpc$^{-2}$,
$\beta = 1.4\pm0.15$ , $N_\perp^{\rm c} = 1.25\times 10^{20}$~cm$^{-2}$
and log~$N_\perp^{\rm crit}$ = 20.62. 
Using the above equation and the $N$(H~{\sc i}) distribution of our sample we would expect 
to measure in our experiment a surface SFR of $5.5\times 10^{-3}$~M$_\odot$~yr$^{-1}$~kpc$^{-2}$.
Assuming DLAs have a radius of 10~kpc gives an expectation of  $1.7\pm0.4$~M$_\odot$~yr$^{-1}$. 
This is higher than our  1.2~M$_\odot$~yr$^{-1}$ 3$\sigma$ upper 
limit derived in the previous Section. 
This suggests that star formation in DLAs is less efficient compared to what is seen in the H~{\sc i} 
disks of nearby galaxies. This is consistent with the conclusions by 
\citet[][]{Wolfe06}.
In the local universe such low star formation efficiencies are seen either
in the outer regions of galaxies or in dwarf galaxies \citep{Roychowdhury09,Bigiel10}.

Using \CIIs\ absorption lines, \citet{Wolfe03a} have 
estimated the SFR per unit area, 
$\left < \dot{\psi_{\star}}\right> = 10^{-2.19}$ and $10^{-1.32}$ $
\rm{M_{\odot}~yr^{-1}~kpc^{-2}}$ for CNM (cold neutral medium) and WNM 
(warm neutral medium) models, respectively. 
%
%
The pure WNM solution 
will require an unphysically low escape fraction ($\le 0.004$) of \lya\ photons
in order for the model to comply with our upper limit.
As DLAs have very little dust content, such low \lya\ escape fraction is unlikely. 
In the case of the CNM solution, for the upper limit we observe to be consistent 
with the \CIIs\ measurements, 
the Ly$\alpha$ escape fraction $f_{\rm esc}^{\rm Ly\alpha}$ should be smaller than 0.03.
This value, although low, is similar to that estimated in LBGs \citep{Hayes10}. 
Note that the high-cool population of C~{\sc ii}$^*$ as defined
by \citet{Wolfe08} will also require unphysically low values of
$f_{\rm esc}^{\rm Ly\alpha}$. In any case, it can be concluded that in situ star 
formation is low in DLAs.
%

Using deep long-slit spectroscopy of an empty field,
\citet{Rauch08} have detected extended and weak \lya\ emitters at 2.67$\le z \le$3.75 
Based on the projected size (a median radius of
0.99'' corresponding to a physical size of 7.7 kpc) and a volume density
of 3$\times 10^{-2} h^3_{70} {\rm Mpc}^{-3}$,
they found that the number per unit redshift of these emitters 
is consistent with that of DLAs.
%
The individual \lya\ flux in these objects
range from 1.5 to 15$\times 10^{-18}$  erg s$^{-1}$ cm$^{-2}$ with 
mean and median fluxes of, respectively, 3.7 and 3$\times$10$^{-18}$~erg~s$^{-1}$~cm$^{-2}$.
If typical DLAs were associated with these 
objects (with sizes less than the projected size of the SDSS fibre), 
then we should have detected \lya\ emission in the stacked spectrum.
%
%
%
Thus the analysis presented here does not support the idea that
the  unresolved or centrally dominated faint \lya\ emitters found by  
\citet{Rauch08} are the host-galaxies of DLAs with log $N$(H~{\sc i}) $\ge$ 20.62.
However, our analysis can not rule out DLAs being similar to their
amorphous extended class of objects with sizes $\ge 10$ kpc.

In summary, there are strong evidences for DLAs to be associated with regions of star-formation:
(i) the median metallicity of DLAs is of the order of 10$^{-1.5}$ solar \citep{Noterdaeme08}
and is always larger than 10$^{-3}$~solar \citep{Penprase10};
(ii) we confirm the findings by Fox et al. (2007) that strong O~{\sc vi} absorption is associated with
DLAs with a velocity spread ($\rm{FWHM}$~$\sim$~250~km~s$^{-1}$) about twice larger
than the velocity spread of the low-ionization gas suggesting bulk-motions in the hot phase
due to kinematical input from supernovae and/or gravitational effects;
(iii) C~{\sc ii}$^*$ is detected in  almost half of the DLAs and can only be explained
by energy input from star-formation activity (Wolfe et al. 2008).
However, we show here that the mean \lya\ emission within 12 kpc from DLAs is smaller than the
emission in most of the weak \lya\ emitters detected by Rauch et al. (2008) and
is consistent with very low in situ star formation rates.
Therefore, the DLA phase could be spread far away from the center of the
star forming regions. The hot phase associated with DLAs could then be the relics of past 
wind episods in the star forming regions.

\section*{acknowledgement}
We acknowledge the use of SDSS spectra from the archive (http://www.sdss.org/). RS and PPJ gratefully 
acknowledge the support from the IFCPAR.
PN is supported by a CONICYT/CNRS fellowship.
\def\aj{AJ}%
\def\actaa{Acta Astron.}%
\def\araa{ARA\&A}%
\def\apj{ApJ}%
\def\apjl{ApJ}%
\def\apjs{ApJS}%
\def\ao{Appl.~Opt.}%
\def\apss{Ap\&SS}%
\def\aap{A\&A}%
\def\aapr{A\&A~Rev.}%
\def\aaps{A\&AS}%
\def\azh{AZh}%
\def\baas{BAAS}%
\def\bac{Bull. astr. Inst. Czechosl.}%
\def\caa{Chinese Astron. Astrophys.}%
\def\cjaa{Chinese J. Astron. Astrophys.}%
\def\icarus{Icarus}%
\def\jcap{J. Cosmology Astropart. Phys.}%
\def\jrasc{JRASC}%
\def\mnras{MNRAS}%
\def\memras{MmRAS}%
\def\na{New A}%
\def\nar{New A Rev.}%
\def\pasa{PASA}%
\def\pra{Phys.~Rev.~A}%
\def\prb{Phys.~Rev.~B}%
\def\prc{Phys.~Rev.~C}%
\def\prd{Phys.~Rev.~D}%
\def\pre{Phys.~Rev.~E}%
\def\prl{Phys.~Rev.~Lett.}%
\def\pasp{PASP}%
\def\pasj{PASJ}%
\def\qjras{QJRAS}%
\def\rmxaa{Rev. Mexicana Astron. Astrofis.}%
\def\skytel{S\&T}%
\def\solphys{Sol.~Phys.}%
\def\sovast{Soviet~Ast.}%
\def\ssr{Space~Sci.~Rev.}%
\def\zap{ZAp}%
\def\nat{Nature}%
\def\iaucirc{IAU~Circ.}%
\def\aplett{Astrophys.~Lett.}%
\def\apspr{Astrophys.~Space~Phys.~Res.}%
\def\bain{Bull.~Astron.~Inst.~Netherlands}%
\def\fcp{Fund.~Cosmic~Phys.}%
\def\gca{Geochim.~Cosmochim.~Acta}%
\def\grl{Geophys.~Res.~Lett.}%
\def\jcp{J.~Chem.~Phys.}%
\def\jgr{J.~Geophys.~Res.}%
\def\jqsrt{J.~Quant.~Spec.~Radiat.~Transf.}%
\def\memsai{Mem.~Soc.~Astron.~Italiana}%
\def\nphysa{Nucl.~Phys.~A}%
\def\physrep{Phys.~Rep.}%
\def\physscr{Phys.~Scr}%
\def\planss{Planet.~Space~Sci.}%
\def\procspie{Proc.~SPIE}%
\let\astap=\aap
\let\apjlett=\apjl
\let\apjsupp=\apjs
\let\applopt=\ao
\bibliographystyle{mn}
\bibliography{mybib}

\end{document}